# Determination of Nanoscale Mechanical Properties of Polymers via Plasmonic Nanoantennas

*Hilario D. Boggiano,[†] Rodrigo Berté,[‡] Alberto F. Scarpettini,[§,⊥] Emiliano Cortés,[‡] Stefan A. Maier,\*[‡,‖] and Andrea V. Bragas\*[†]*

[†]Departamento de Física, FCEN, IFIBA CONICET, Universidad de Buenos Aires, Intendente Güiraldes 2160, C1428EGA Buenos Aires, Argentina

[‡]Chair in Hybrid Nanosystems, Nanoinstitute Munich, Faculty of Physics, Ludwig-Maximilians-Universität München, 80539 München, Germany

[§]Grupo de Fotónica Aplicada, Facultad Regional Delta, Universidad Tecnológica Nacional, 2804 Campana, Argentina

[⊥]Consejo Nacional de Investigaciones Científicas y Técnicas (CONICET), 1425 Buenos Aires, Argentina

[‖]Department of Physics, Imperial College London, London SW7 2AZ, UK

ABSTRACT: Nanotechnology and the consequent emergence of miniaturized devices are driving the need to improve our understanding of the mechanical properties of a myriad of materials. Here we focus on amorphous polymeric materials and introduce a new way to determine the nanoscale mechanical response of polymeric thin films in the GHz range, using ultrafast optical means. Coupling of the films to plasmonic nanoantennas excited at their vibrational eigenfrequencies allows the extraction of the values of the mechanical moduli as well as the estimation of the glass transition temperature via time-domain measurements, here demonstrated for PMMA films. This nanoscale method can be extended to the determination of mechanical and elastic properties of a wide range of spatially strongly confined materials.

Polymers form a constitutive part of functional coatings, sensors, microfluidic or optoelectronic devices but also, they are at the core of many nano- and microfabrication processes. Successful implementation in these real-world applications requires detailed knowledge, in addition to careful optimization, of a wide range of mechanical properties adapted to specific uses. Regarding thin films and other confined geometries, the mechanical characterization should be performed on spatial scales akin to the physical sizes of the nanostructures, whereas dynamic mechanical behavior needs to be tested in-operando conditions at frequencies of the order of GHz, in accordance with the operating frequencies of many current devices and potential applications.[1–3]

Conventional bulk-related methods for mechanical assays are not usually sensitive to the measurement of elastic properties in thin films. Nanoindentation is commonly employed as a successful method to obtain the mechanical moduli of thin films made of hard materials.[4] However, it has some limitations when treating soft materials, for which it tends to overestimate the values due to several reasons such as the stiffness of the substrate, the effect of the tip and the viscoelastic nature of the material.[5] There are other methods to measure Young's moduli of thin films based on ellipsometry,[6] strain-induced elastic buckling instability,[7] capacitance changes under hydrostatic pressure,[8] and Brillouin light scattering,[9,10] of successful implementation but with scarce possibility of studying point local effects on the nanometer scale. Besides, except for the latter, all these methods lack on the characterization of the dynamical response in the GHz range, for which high-frequency electroacoustic or photoacoustic experimental designs have demonstrated to be much better suited. For instance, it is possible to study the elastic properties of thin films by characterizing the speed of propagation of surface acoustic waves (SAW) electrically,[11] or optically generated,[12] or by using a metal layer transducer deposited on top of the sample and excited with a laser pulse, generating a strain wave that propagates towards the sample.[13–15] This last technique constitutes a big step towards hypersound nanoimaging and nanotomography, but with up-to-now unresolved important limitations when examining local mechanical properties.

The ability and efficiency of plasmonic nanoantennas excited via an ultrafast light pulse for the generation of coherent acoustic phonons have been comprehensively demonstrated.[16–19] These initial studies have promoted an increasing interest in the construction of tailored individual plasmonic nanoresonators as sources and detectors of hypersound.[20–23] In the present work we propose a new method for the measurement of local mechanical properties of polymeric thin films using plasmonic nanoantennas vibrating at GHz frequencies. The use of plasmonic nanoantennas as sources of hypersound may have advantages compared with its film transducer counterpart. These nanoobjects can be embedded and buried in the materials under study, grown or deposited at specific sites over surfaces and/or synthetized with surface molecular modifiers for optimal mechanical contact. Also, they may act as sources or as detectors of minute mechanical vibrations, well below the spatial resolution of diffraction-limited optics.[21]

After the excitation of a plasmonic nanoantenna with pulsed light at the appropriate wavelength (usually its dipolar plasmon resonance, or an interband transition), an excited electronic population is produced in the metal followed by thermalization and lattice heating. In addition to mere heat, this energy transfer cascade leads to the coherent excitation of normal modes of mechanical oscillation compatible with a symmetric strain profile of the plasmonic nanostructure.[20] The induced field of coherent acoustic phonons in turn modulates the optical properties of the nanoantenna such as its reflection or transmission. This dynamic change can be read by a second probe pulse, for which there will be maximum sensitivity at wavelengths near the localized dipolar surface plasmonic resonance (LSPR) of the antenna, as shown in Figure 1a,b. For the Au nanorods antennas (GNRs) used in this work, with dimensions shown in Figure 1a, vibrational modes lie in the GHz range. A detailed description of the expected oscillation behavior can be obtained via numerical solutions of the elastic problem, including the interaction with a surrounding polymeric material and the substrate. On the other hand, experiments will detect vibrations with the greatest amplitudes of motion in any of the dimensions of the rod, among which the extensional mode, which implies mainly longitudinal displacements, is by far the most dominant for this geometry. Therefore, in the present paper we study in detail the extensional mode generated in several individual GNRs, with the aim to measure the local (and dynamic) mechanical response of a surrounding polymeric material in the GHz range. Indeed, using a pump-probe setup and comparing the experiments with numerical simulations we show how we can read out the polymer's elastic parameters from the frequencies and quality factors of the vibrational plasmonic oscillators. Additionally, frequency shifts of the extensional mode that resemble the characteristic polymer glass transition curves are also observed by controlling changes of the surrounding temperature with the pump intensity.

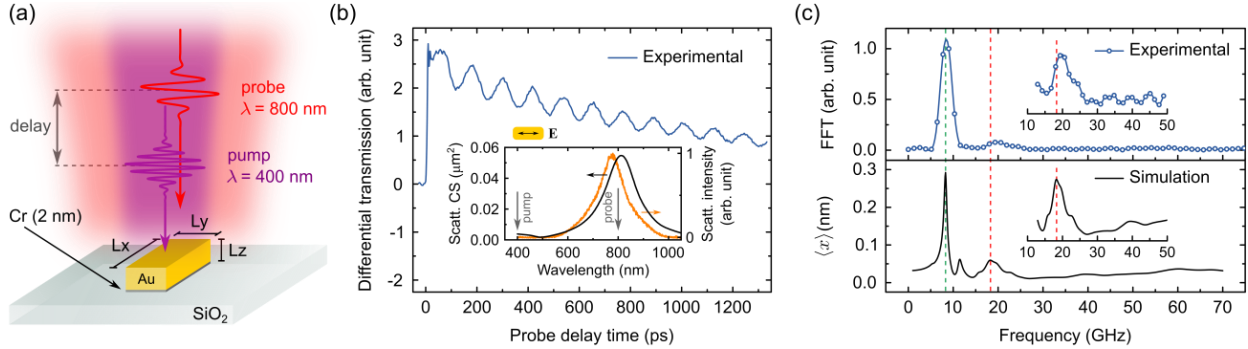

**Figure 1.** Generation of coherent acoustic phonons in air-surrounded gold nanoantennas. (a) Sketch of the non-degenerate pump-probe setup for coherent acoustic phonon detection. Gold nanorod (GNR) nominal dimensions Lx, Ly and Lz are 140 × 60 × 35 nm, respectively. (b) Experimental differential probe transmission of a single GNR antenna as a function of the probe delay time. The inset shows the experimental scattering intensity and the simulated scattering cross section of the GNR with the plasmonic resonance at about 800 nm. (c) Top: FFT magnitude of the signal show in (b). Bottom: simulated eigenmode spectrum of the average amplitude for the displacement in *x*-direction in the rod spatial domain as a function of the mechanical oscillation frequency. In both panels, green (red) dashed line indicates the mechanical resonance frequency of the extensional (breathing-like) mode obtained from FEM simulations. Insets show a zoom of the same data, where the peak corresponding to the breathing-like mode is enlarged observed.

Using electron-beam lithography, we have fabricated GNRs attached to a quartz substrate, via a two-nanometer thick Cr adhesion layer.[23] A representative scattering spectrum is shown in the inset of Figure 1b. Experiments were carried out on individual GNRs using a non-degenerate pump-probe setup, as sketched in Figure 1a. The second harmonic (400 nm) of a Ti:sapphire laser with repetition rate $f_r$ = 95 MHz, 100 fs pulsewidth and modulated with an acousto-optic modulator acts as the pump, whereas a delayed probe is set at 800 nm. Both beams are focused using a high numerical aperture microscope objective onto a single GNR with a pump (probe) spot size diameter of 2.5 µm (1 µm). The signal is collected in transmission after optical filtering and sent to a lock-in amplifier. Additionally, dark field images of the sample can be taken in the same setup, using grazing illumination by deflecting the same beam that is used as a probe pulse. When the coherent phonon field is present, the plasmon frequency modulation results in changes in the transmission of a delayed probe pulse which measures the time-resolved signal as the one shown in Figure 1b. Experimental errors and the fitting procedure used throughout this work, by using singular value decomposition methods,[24] cast experimental errors in the determination of the frequencies as low as a half percent on average. The fast Fourier transform (FFT) in Figure 1c reveals that two modes were detected in this case, one is an extensional (lower frequency) and the other, much weaker, a breathing-like mode (higher frequency), as previously demonstrated for this configuration.[20] To further support the observations shown all along this work, we performed finite-element method (FEM) calculations by solving Navier's equation using the commercially available software COMSOL Multiphysics to evaluate the mechanical responses of nanostructures. Details of the calculations are given in the Supporting Information S1. As it can be seen from the data shown in Figure 1c, the correspondence between the simulated mechanical modes and the experiment is quite good, although in general there would be small discrepancies due to the size dispersion of the sample and imperfect mechanical contact with the substrate.

In what follows, we will show how the mechanical parameters of the nanoresonator change when surrounded by a polymer, focusing on the extensional mode. For that, we spin-coated the sample at 3000 rpm for 1 minute with Poly(methyl methacrylate) PMMA (MicroChem, $M_w$ = 950 kg mol$^{-1}$) dissolved in anisole and let the solvent evaporate at room temperature for one day before doing any measurements, resulting in a 0.3 μm thick film. We have explored the behavior of many different individual GNRs before and after covering them with the PMMA film, i.e., we measure frequency, amplitude and quality factor for the same air-surrounded and PMMA-surrounded nanoantennas. This is sketched in Figure 2a.

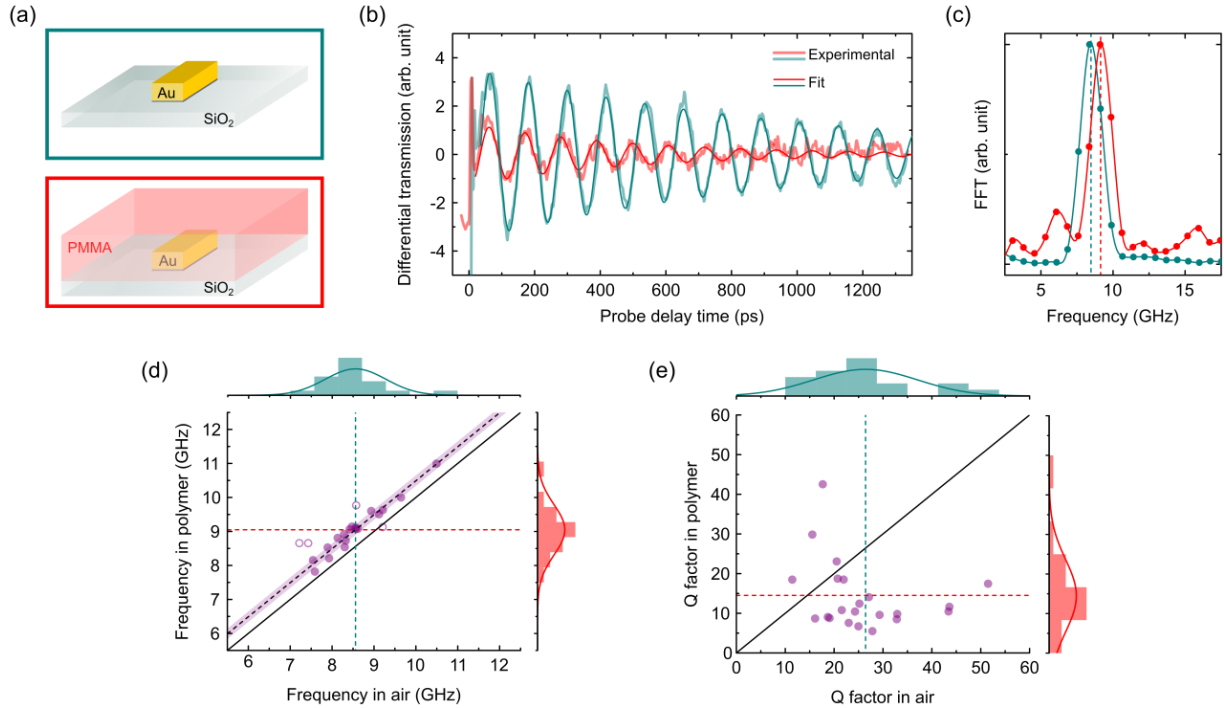

**Figure 2.** Comparison of air-surrounded (cyan) and PMMA-surrounded (red) nanoantennas. The systems are schematically represented in (a). (b) Differential probe transmission of the same antenna in air and when covered with the polymer, showing the fits which consist in a sum of exponentially damped sinusoids (exponential decay subtracted) where each term involves four parameters: amplitude, frequency *f*, phase and an exponential decay time τ. (c) FFT magnitude for experimental signals showed in (b). (d,e) Frequency and *Q* factor ($Q = \pi f \tau$) measurements on several individual nanoantennas, corresponding to the extensional mode of oscillation. Measurements were performed first with antennas surrounded by air, and then coated with the polymer. The solid black line is the identity, and dashed lines indicates the mean values in air (cyan) $\bar{f}^{\text{air}} = (8.5 \pm 0.7)$ GHz, in polymer (red) $\bar{f}^{\text{PMMA}} = (9.0 \pm 0.7)$ GHz, and the mean increment (black) $\overline{\Delta f} = (0.5 \pm 0.2)$ GHz  ($\Delta f_i = f_i^{\text{PMMA}} - f_i^{\text{air}}$) where the purple shadowed area shows the dispersion (standard deviation). Outliers are marked as hollow circles and have not been considered in statistics. Histograms represent the distribution of experimental data for each axis.

The first observation that can be made by comparing both kinds of systems is that the frequency is higher when the nanoantenna is covered by PMMA, as shown in the example of Figure 2b,c. In the same way as the substrate adds an extra restoring force compared with the (ideal) picture of an isolated vibrating nanoantenna, thus increasing the frequency of its modes,[23] the polymer environment adds up a restoring force as well. This contributes to the vibrational frequency in a similar functional way as the Au Young's modulus does and could be in principle considered as an effective internal elastic modulus, higher than the original one. Indeed, in Figure 2d, measurements of the frequencies $f_i^{\text{air}}$ and $f_i^{\text{PMMA}}$ over more than 20 antennas show clearly that the frequency increases when the antenna is PMMA-surrounded (i.e. see black line in Figure 2d, the identity, to clearly visualize this effect). It is worth to note that the dispersion of values around the averages (dashed lines) is much lower for $\Delta f_i = f_i^{\text{PMMA}} - f_i^{\text{air}}$ than for each frequency separately, which makes sense because, if $f_i^{\text{air}}$ and $\Delta f_i$ are independent (i.e. no correlations), then it is satisfied for the variances that $\sigma^2(f^{\text{PMMA}}) = \sigma^2(f^{\text{air}}) + \sigma^2(\Delta f)$. This result indicates that all the indetermination coming from the size dispersion and mechanical contact disappears with this subtraction and therefore $\Delta f_i$ gives us only the information about the local mechanical properties of the polymer.

Figure 2e also shows how the $Q$ factors of the nanoresonators change. Unlike the case of air, the polymeric environment offers much better conditions for energy dissipation from the GNRs. For air, there is virtually no acoustic radiation escaping through the upper and lateral surfaces of the GNR given the huge air-Au impedance mismatch, and the only channel is across the small part of the GNR in contact with the quartz substrate.[21] However, when the GNR is surrounded with the polymer, the whole surface of the nanorod is in mechanical contact with the environment, through which it will radiate sound and heat. As a result, the $Q$ factor lowers when the nanoantenna is PMMA-surrounded. This can be seen from Figure 2e, where despite few outliers, the cloud of points is located below the identity line.

In order to extract quantitative values for the mechanical moduli of PMMA we compare extensive numerical simulations of the polymer-induced frequency shifts to our experimental results. Considering the average dimensions given in Figure 1a and values in Table 1,[25,26] a frequency of $f_{\text{sim.}}^{\text{air}} = 8.3$ GHz is obtained from FEM simulations for the extensional mode (Figure 3b, black curve). Additionally, the simulated spectral response, shown as the average $x$-displacement amplitude in the nanoparticle, is obtained in Figure 3b for different possible values of the polymer shear modulus, $G$. It is clearly seen how $f_{\text{sim.}}^{\text{PMMA}}$ moves to higher frequencies when $G$ increases. Equivalently, the same shift behavior can be interpreted as an increase in the polymer's Young's modulus ($E$), which for the approximation of an isotropic homogeneous linear elastic material is related to $G$ through the Poisson ratio (ν) via $G = E/[2(1+ ν)]$. The value of ν was set constant in all simulations as given in Table 1.[27,28] Eigenfrequency calculations of the extensional mode frequency $f_{\text{sim.}}^{\text{PMMA}}$ as a function of the moduli ($G$ and $E$) are shown in the inset of Figure 3b. $f_{\text{sim.}}^{\text{PMMA}}$ has a quasi-linear behavior, with a slope almost invariant when calculated for different antenna sizes (within 10% between the smallest and the largest simulated antennas). This behavior allows us to disregard the actual sizes of the individual nanoantennas when extracting local values of $G$ from the frequency shift, $\Delta f = f^{\text{PMMA}} - f^{\text{air}}$, instead of $f^{\text{PMMA}}$. As discussed above, $\Delta f$ would carry the lower dispersion introduced by the method variabilities, so then we compute a linear fit for $G$ vs the FEM-predicted frequency shift shown in Figure 3c. Now, by using the experimental average value $\overline{\Delta f} = 0.5$ GHz marked in a vertical red dashed line in Figure 3c we get the average values of $\bar{G}_{\text{exp.}} = (1.1 \pm 0.3)$ GPa and $\bar{E}_{\text{exp.}} = (3.0 \pm 0.9)$ GPa (horizontal red dashed line) for the shear and elastic moduli of PMMA respectively in a nanometer confined geometry at around 8 GHz.

The shear modulus in the glassy bulk state is reported to be around 1.7 GPa, at 25ºC and 10 Hz.[29] However, it has been shown that the value of the shear modulus of PMMA increases with frequency,[30] reaching values as high as 2.5 GPa at 200 kHz for bulk, although it drops to 1 GPa for a 12 μm thickness film at 9 MHz.[31] Values of 3.2 GPa at 4.2 K and 0.1 Hz,[32] and 5 GPa at 77 K and 500 Hz,[33] have also been reported. As far as we know, there is only one paper in the GHz range reporting a value of 1.9 GPa at 0.5 GHz for a multilayered film.[34] This spread in values found in the literature are due to the different excitation frequencies, experimental methods, film thickness, temperature and molecular weight, which makes difficult a simple and direct comparison. A similar situation features for de Young's modulus of PMMA film, for which it is reported 2.34 GPa using microindentation methods,[35] and ~3 GPa for films above 40 nm thickness using the polymer buckling technique,[36] which nonetheless compare fairly well with the values we obtain.

**Table 1.** Reference values for properties of materials used in numerical simulations.

| Material | Young's modulus [GPa] | Poisson's ratio | Mass density [kg m$^{-3}$] | Coefficient of thermal expansion [K$^{-1}$] | Specific heat capacity [J kg$^{-1}$ K$^{-1}$] | Thermal conductivity [W m$^{-1}$ K$^{-1}$] |
|---|---|---|---|---|---|---|
| Au | 78 | 0.44 | 19300 | 14.2 × 10$^{-6}$ | 129 | 315 |
| Cr | 279 | 0.21 | 7150 | | 449 | 93.9 |
| Quartz | 73 | 0.17 | 2200 | | 703 | 1.4 |
| PMMA | 1-7 | 0.37 | 1180 | | 1466 | 0.2 |

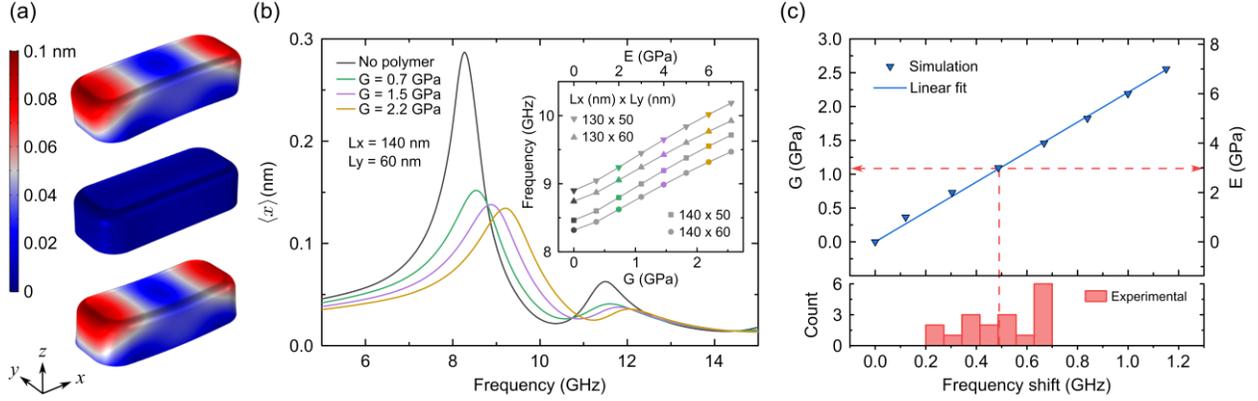

**Figure 3.** (a) View of the 3D FEM simulation of the extensional mode for a 10 K increase in the lattice temperature. A 50× scale factor was applied to the deformation in order to highlight the displacement map. Top and bottom snapshots are taken at the phases corresponding to the highest expansion and contraction respectively. (b) Simulated frequency-domain spectrum for a 140 × 60 × 35 nm GNR in air and surrounded by the polymer with different values of the shear modulus $G$. The inset shows eigenfrequency calculations of the dependence of the extensional mode frequency with $G$ for different antenna sizes. (c) Relation between shear modulus and frequency shift extracted from simulations with corresponding linear fit used to estimate the $G$ and $E$ values from the average of the experimental frequency shift values (red dashed line). Experimental values are shown in the bottom panel.

Additionally, $G$ is a complex number, in which the real part is the storage modulus and the imaginary part is the loss modulus (at least one order of magnitude lower for the glassy state).[37] For this polymeric viscoelastic environment, the acoustic energy generated by the GNR may be both stored and dissipated in the PMMA, so in principle both contributions would need to be considered. However, the imaginary part is so small that the oscillator parameters will not be much sensitive to it according to the analytical model presented in ref. 38, and the simulations shown in Supporting Information S3.

Nevertheless, this method is also capable of determining the actual local value for each position of the antennas besides the average values if $\Delta f$ is known. The histogram with values of $G$ extracted from the experimental frequencies is shown at the bottom panel of Figure 3c. There is a distribution of values around $\bar{G}_{\text{exp.}}$, with a standard deviation which is solely due to differences in the mechanical properties of the nanometric environment around the GNR, that could be explained by different polymer-GNR attachment, morphologies and density distributions. Now we turn to evaluate the dynamic properties of the polymeric environment due to light-induced heating of the nanoresonators.

It is well known that plasmonic nanoantennas are efficient light to heat converters. Indeed, after the optical excitation, the GNR absorbs the energy of the pulse and reaches electronic-lattice thermalization in a few picoseconds timescale, much faster than the time-scale of hundreds of picosecond up to ten nanoseconds for the GNR cooling due to heat transfer to the environment.[16,39,40] So far, the measurements presented were taken with small incident power in order not to significantly heat the nanoantenna and the polymer, thus avoiding variations in the elastic properties of the latter. In what follows we present measurements where the incident power is raised in order to get a controlled increment of the temperature of each individual nanoantenna. The maximum lattice temperature depends on the GNR absorption coefficient and the incident irradiance, while the dissipation time depends also on the thermal properties of the surrounding media. Nevertheless, absorbed energy can be dissipated to the environment before the next pulse arrives –if the time between pulses is longer than heat dissipation rates–, see calculations presented in Figure S4.1 and Figure S4.2 in the Supporting Information, thus avoiding heat storage and the concomitant temperature raising of the GNR with time.

The temperature can be estimated when the GNR absorbs one pulse energy as $T = \frac{\sigma_{\text{abs}} \langle I \rangle}{f_r V_{\text{NR}} \rho_{\text{Au}} c_{\text{Au}}} + T_r$,[39,41] where the room temperature $T_r \approx 23°C$, $\sigma_{\text{abs}}$ is the absorption cross section, $\langle I \rangle$ is the average irradiance, $f_r$ is the laser repetition rate, $V_{\text{NR}}$ is the GNR volume, $\rho_{\text{Au}}$ and $c_{Au}$ are the gold density and heat capacity at constant pressure, respectively. As can be seen, the maximum temperature increment does not depend on the thermal properties of the environment, because the peak temperature of the nanoparticle is reached before the dissipation to the surroundings begins.[16,40]

Under the hypothesis that the system thermalizes before the arrival of the next pulse, any heat integration effect can be neglected and therefore only the effect of temperature increment due to the absorption of the pump beam is observed in the experiments. Also, our single particle excitation and readout experiments prevent collective heating effects, previously reported for plasmonic systems.[42,43] This is confirmed by the full numerical calculations for the time evolution of the temperature shown in Supporting Information S4 using the absorption cross section at 400 nm (at the incident polarization parallel to the main axis of the nanorod) and the parameters shown in Table 1. From these calculations, we can see that the equation for the temperature described above gives a good estimation of the maximum temperature, although overestimates the value of the average temperature over the whole length of the experiment (~1.3 ns) (see Figure S4.1). On the other hand, any small residual heat accumulated after each pulse period and the presence of an interface thermal resistance, both slightly raise the temperature. Without loss of generality, we thus use this calculation in the rest of the work knowing that the calculated temperatures serve as a rough estimation.

Figure 4 shows the results of measurements on three different individual GNR, covered by two different polymeric films, for which the incident power has been progressively increased. For the annealed PMMA-film we bake the sample at a rate of 10 K/min until reaching a maximum temperature of 140ºC, at which the sample was heated for 5 min. In all the cases, a clear monotonous decrease in the extensional mode frequency associated with the light-induced increase in temperature is seen in different temperature ranges for the different samples, as shown in the shadowed areas of Figure 4. Indeed, as the temperature increases, the nanoresonator senses the surrounding polymer alterations, leading to lower frequency shifts of the extensional mode relative to the GNR in an air environment. This means that the polymer softens (lower $G$) as the pump intensity becomes higher (higher light-induced temperatures), according to the analysis presented in Figure 3c. Moreover, the observed behavior in Figure 4 resembles the glass transition of the PMMA, from which a glass transition temperature, $T_g$, can be taken. This transition in polymers is defined as a gradual and reversible process, in which the polymer passes from a glassy to a rubbery state as the temperature rises. The glass transition is mainly due to the change in the mobility of the polymer chains, as it experiences changes in intermolecular interactions and in the structure and orientation of the chains.[44,45] As the temperature rises the chains are able to slide over each other, dramatically decreasing the rigidity of the material; thus, the knowledge of $T_g$ defines and allows one to predict the mechanical behavior of the polymer and consequently their potential applications in mesoscopic devices, in this case, in the GHz regime. In order to extract the values of $T_g$ we employ the Dupaix-Boyce constitutive model,[45] which describes the stress-strain behavior of an amorphous polymer with temperature through the glass transition, connecting the glassy and the rubbery states (refer to Supporting Information S5 for details). This model represents the strong dependence of the elastic moduli with the temperature.

Figure 4 shows that the mechanical response of the antennas surrounded by the annealed or non-annealed polymers is substantially different, regarding the glass transition temperature. From the experimental data of Figure 4 and the corresponding fits by the Dupaix-Boyce model, we have obtained an average $T_g$ of 39ºC for the non-annealed polymer film, and a value of 74ºC for the annealed one, which might indicate a strong influence of any residual solvent trapped in the PMMA layers of the former samples, yielding a reduction in $T_g$. Nevertheless, a great variety of values of $T_g$ for PMMA obtained from different physical methods and taken under different conditions are reported in the literature. An overview of this huge dispersion in values can be found in ref. 46, where values of $T_g$ from 30ºC to 190ºC have collected over 20 different test methods. It should be considered that this dispersion of values is partly due to diverse influencing factors that control the value of $T_g$ as the molecular weight distribution of the sample, the tacticity, presence of plasticizers, geometric confinement, heating rate and the scale factor. In order to extract the elastic parameters of the glass transition of polymeric film, a numerical analysis was employed for the annealed sample (Supporting Information S5) showing excellent agreement with experimental results.

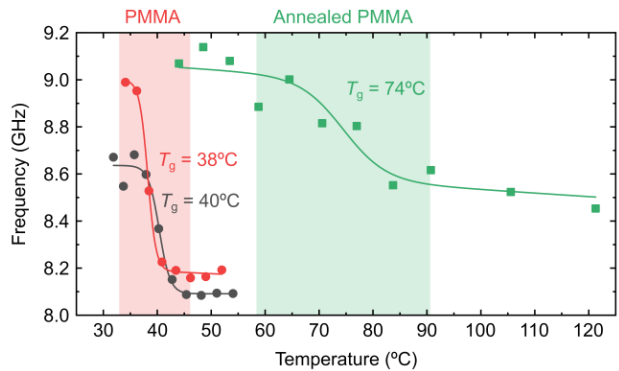

**Figure 4.** Frequency measurements as a function of the temperature for three different GNRs embedded in annealed and non-annealed PMMA films. Fits with the Dupaix-Boyce constitutive model for polymers across the glass transition temperature are shown as solid lines.

To conclude, we introduced a novel optical method for nanoscale measurements of the mechanical moduli of polymers at GHz frequencies. The method relies on the optical readout of the mechanical vibration of individual plasmonic nanoantennas. FEM simulations enable quantitative read-out of the mechanical moduli thorough the knowledge of the oscillation frequencies, allowing the determination of the average and even the local values. Hence, our method can be employed as a nanometrological probe of mechanical properties of polymeric thin films. Additionally, it has been shown that by increasing the incoming pump power it is possible to induce an extensional mode frequency behavior that resembles a glass transition on different polymeric films, showing that the given method is able to sense the light-induced local mechanical changes of the environment. The measurements presented reveal the importance of counting on local methods to evaluate the mechanical performance of polymers, which are widely used in current technological applications.

ASSOCIATED CONTENT

**Supporting Information**. Supporting Information Available: Mechanical numerical simulations, optical response of gold nanorod in air and surrounded by the polymer, complex part of the shear modulus $G$, temperature calculations-simulations, numerical calculations employing the Dupaix-Boyce model. This material is available free of charge via the internet at http://pubs.acs.org.

AUTHOR INFORMATION

**Corresponding Author**

*E-mail: bragas@df.uba.ar

*E-mail: stefan.maier@physik.uni-muenchen.de

**Author Contributions**

H.D.B. and R.B. contributed equally to this work.

**Notes**

The authors declare no competing financial interest.

ACKNOWLEDGMENT

The authors acknowledge Mohsen Rahmani for fabrication of the nanoantennas. This work was partially supported by PICT 2017-2534, PIP 112 201301 00619, UBACyT Proyecto 200020170100432BA, PID-UTI4836 and by the Deutsche Forschungsgemeinschaft (Germany´s Excellence Strategy – EXC 2089/1 – 390776260) and European Commission, ERC-802989 (Catalight).

REFERENCES

(1) Chen, G.; Zhao, X.; Wang, X.; Jin, H.; Li, S.; Dong, S.; Flewitt, A. J.; Milne, W. I.; Luo, J. K. Film bulk acoustic resonators integrated on arbitrary substrates using a polymer support layer. *Sci. Rep.* **2015**, *5*, 9510.

(2) Travagliati, M.; Nardi, D.; Giannetti, C.; Gusev, V.; Pingue, P.; Piazza, V.; Ferrini, G.; Banfi, F. Interface nano-confined acoustic waves in polymeric surface phononic crystals. *Appl. Phys. Lett.* **2015**, *106*, 021906.

(3) Kumar, A.; Thachil, G.; Dutta, S. Ultra high frequency acoustic wave propagation in fully polymer based surface acoustic wave device. *Sens. Actuators, A.* **2019**, *292*, 52–59.

(4) Gouldstone, A.; Chollacoop, N.; Dao, M.; Li, J.; Minor, A. M.; Shen, Y. L. Indentation across size scales and disciplines: Recent developments in experimentation and modeling. *Acta Mater.* **2007**, *55*, 4015–4039.

(5) Mogilnikov, K. P.; Baklanov, M. R. Determination of Young's Modulus of Porous Low-k Films by Ellipsometric Porosimetry. *Electrochem. Solid-State Lett.* **2002**, *5*, 29–31.

(6) Boissiere, C.; Grosso, D.; Lepoutre, S.; Nicole, L.; Bruneau, A. B.; Sanchez, C. Porosity and Mechanical Properties of Mesoporous Thin Films Assessed by Environmental Ellipsometric Porosimetry. *Langmuir* **2005**, *21*, 12362–12371.

(7) Stafford, C. M.; Harrison, C.; Beers, K. L.; Karim, A.; Amis, E. J.; Vanlandingham, M. R.; Kim, H. C.; Volksen, W.; Miller, R. D.; Simonyi, E. E. A buckling-based metrology for measuring the elastic moduli of polymeric thin films. *Nat. Mater.* **2004**, *3*, 545–550.

(8) van Soestbergen, M.; Ernst, L. J.; Jansen, K. M. B.; van Driel, W.D. Measuring the through-plane elastic modulus of thin polymer films in situ. *Microelectron. Reliab.* **2007**, *47*, 1983–1988.

(9) Forrest, J. A.; Dalnoki-Veress, K.; Dutcher, J. R. Brillouin light scattering studies of the mechanical properties of thin freely standing polystyrene films. *Phys. Rev. E* **1998**, *58*, 6109–6114.

(10) Hartschuh, R. D.; Kisliuk, A.; Novikov, V.; Sokolov, A. P.; Heyliger, P. R.; Flannery, C. M.; Johnson, W. L.; Soles, C. L; Wu, W.-L. Acoustic modes and elastic properties of polymeric nanostructures. *Appl. Phys. Lett.* **2005**, *87*, 173121.

(11) Tetelin, A.; Blanc, L.; Tortissier, G.; Dejous, C.; Rebière, D.; Boissière, C. Guided SH-SAW Characterization of Elasticity Variations of Mesoporous $TiO_2$ Sensitive Films during Humidity Sorption. *Proc. IEEE Sens.* **2010**, 2136–2140.

(12) Flannery, C. M.; Murray, C.; Streiter, I.; Schulz, S. E. Characterization of thin-film aerogel porosity and stiffness with laser-generated surface acoustic waves. *Thin Solid Films* **2001**, *388*, 1–4.

(13) Hettich, M.; Jacob, K.; Ristow, O.; Schubert, M.; Bruchhausen, A.; Gusev, V.; Dekorsy, T. Viscoelastic properties and efficient acoustic damping in confined polymer nano-layers at GHz frequencies. *Sci. Rep.* **2016**, *6*, 33471.

(14) Mechri, C.; Ruello, P.; Breteau, J. M.; Baklanov, M. R.; Verdonck, P.; Gusev, V. Depth-profiling of elastic inhomogeneities in transparent nanoporous low- k materials by picosecond ultrasonic interferometry. *Appl. Phys. Lett.* **2009**, *95*, 091907.

(15) Lomonosov, A. M.; Ayouch, A.; Ruello, P.; Vaudel, G.; Baklanov, M. R.; Verdonck, P.; Zhao, L.; Gusev, V. E. Nanoscale Noncontact Subsurface Investigations of Mechanical and Optical Properties of Nanoporous Low-k Material Thin Film. *ACS Nano* **2012**, *6*, 1410–1415.

(16) Crut, A.; Maioli, P.; Del Fatti, N.; Vallée, F. Acoustic vibrations of metal nano-objects: Time-domain investigations. *Phys. Rep.* **2015**, *549*, 1–43.

(17) Hartland, G. V. Coherent Excitation of Vibrational Modes in Metallic Nanoparticles. *Annu. Rev. Phys. Chem.* **2006**, *57*, 403–430.

(18) Yu, K.; Zijlstra, P.; Sader, J. E.; Xu, Q. H.; Orrit, M. Damping of Acoustic Vibrations of Immobilized Single Gold Nanorods in Different Environments. *Nano Lett.* **2013**, *13*, 2710–2716.

(19) Jais, P. M.; Murray, D. B.; Merlin, R.; Bragas, A. V. Metal Nanoparticle Ensembles: Tunable Laser Pulses Distinguish Monomer from Dimer Vibrations. *Nano Lett.* **2011**, *11*, 3685–3689.

(20) Della Picca, F.; Berte, R.; Rahmani, M.; Albella, P.; Bujjamer, J. M.; Poblet, M.; Cortés, E.; Maier, S. A.; Bragas, A. V. Tailored Hypersound Generation in Single Plasmonic Nanoantennas. *Nano Lett.* **2016**, *16*, 1428–1434.

(21) Berte, R.; Della Picca, F.; Poblet, M.; Li, Y.; Cortés, E.; Craster, R. V.; Maier, S. A.; Bragas, A. V. Acoustic Far-Field Hypersonic Surface Wave Detection with Single Plasmonic Nanoantennas. *Phys. Rev. Lett.* **2018**, *121*, 253902.

(22) O'Brien, K.; Lanzillotti-Kimura, N. D.; Rho, J.; Suchowski, H.; Yin, X.; Zhang, X. Ultrafast acousto-plasmonic control and sensing in complex nanostructures. *Nat. Commun.* **2014**, *5*, 4042.

(23) Chang, W. S.; Wen, F.; Chakraborty, D.; Su, M. N.; Zhang, Y.; Shuang, B.; Nordlander, P.; Sader, J. E.; Halas, N. J.; Link, S. Tuning the acoustic frequency of a gold nanodisk through its adhesion layer. *Nat. Commun.* **2015**, *6*, 7022.

(24) Barkhuijsen, H.; de Beer, R.; Bovée, W. M. M. J.; van Ormondt, D. Retrieval of Frequencies, Amplitudes, Damping Factors, and Phases from Time-Domain Signals Using a Linear Least-Squares Procedure. *J. Magn. Reson.* **1985**, *61*, 465–481.

(25) *CRC Handbook of Chemistry and Physics: A Ready-Reference Book of Chemical and Physical Data: 2013-2014, 94th ed.*; Haynes, W. M., Lide, D. R., Bruno, T. J., Eds.; CRC Press: Boca Raton, FL, 2013.

(26) Wu, B.; Heidelberg, A.; Boland, J. J. Mechanical properties of ultrahigh-strength gold nanowires. *Nat. Mater.* **2005**, *4*, 525–529.

(27) Fukuhara, M.; Sampei, A. Low-temperature Elastic Moduli and Internal Dilational and Shear Friction of Polymethyl Methacrylate. *J. Polym. Sci., Part B: Polym. Phys.* **1995**, *33*, 1847–1850.

(28) Abdel-Wahab, A. A.; Ataya, S.; Silberschmidt, V. V. Temperature-dependent mechanical behaviour of PMMA: Experimental analysis and modelling. *Polym. Test.* **2017**, *58*, 86–95.

(29) Wunderlich, W. In *Polymer Handbook*; Brandrup, J.; Immergut, E. H.; Grulke, E. A., Eds.; John Wiley: New York, NY, 1975.

(30) Capodagli, J.; Lakes, R. Isothermal viscoelastic properties of PMMA and LDPE over 11 decades of frequency and time: a test of time-temperature superposition. *Rheol. Acta* **2008**, *47*, 777–786.

(31) Morray, B.; Li, S.; Hossenlopp, J.; Cernosek, R.; Josse, F. PMMA Polymer Film Characterization Using Thickness-Shear Mode (TSM) Quartz Resonator. *Proc. IEEE Int. Freq. Control Symp. PDA Exhib.* **2002**, 294–300.

(32) Sinnott, K. M. Shear Modulus and Internal Friction of Polymethyl Methacrylate and Polyethyl Methacrylate between 4.2 and 100°K. *J. Polym. Sci.* **1959**, *35*, 273–275.

(33) Hoff, E. A. W.; Robinson, D. W.; Willbourn, A. H. Relation between the Structure of Polymers and Their Dynamic Mechanical and Electrical Properties. Part II. Glassy State Mechanical Dispersions in Acrylic Polymers. *J. Polym. Sci.* **1955**, *18*, 161–176.

(34) Saini, G.; Pezeril, T.; Torchinsky, D. H.; Yoon, J.; Kooi, S. E.; Thomas, E. L.; Nelson, K. A. Pulsed laser characterization of multicomponent polymer acoustic and mechanical properties in the sub-GHz regime. *J. Mater. Res.* **2007**, *22*, 719–723.

(35) Amitay-Sadovsky, E.; Wagner, H. D. Evaluation of Young's modulus of polymers from Knoop microindentation tests. *Polymer* **1998**, *39*, 2387–2390.

(36) Stafford, C. M.; Vogt, B. D.; Harrison, C.; Julthongpiput, D.; Huang, R. Elastic Moduli of Ultrathin Amorphous Polymer Films. *Macromolecules* **2006**, *39*, 5095–5099.

(37) Yee, A. F.; Takemori, M. T. Dynamic Bulk and Shear Relaxation in Glassy Polymers. I. Experimental Techniques and Results on PMMA. *J. Polym. Sci., Polym. Phys. Ed.* **1982**, *20*, 205–224.

(38) Wang, L.; Takeda, S.; Liu, C.; Tamai, N. Coherent Acoustic Phonon Dynamics of Gold Nanorods and Nanospheres in a Poly(vinyl alcohol) Matrix and Their Temperature Dependence by Transient Absorption Spectroscopy. *J. Phys. Chem. C* **2014**, *118*, 1674–1681.

(39) Baffou, G.; Rigneault, H. Femtosecond-pulsed optical heating of gold nanoparticles. *Phys. Rev. B* **2011**, *84*, 035415.

(40) Brongersma, M. L.; Halas, N. J.; Nordlander, P. Plasmon-induced hot carrier science and technology. *Nat. Nanotechnol.* **2015**, *10*, 25–34.

(41) Della Picca, F.; Gutiérrez, M. V.; Bragas, A. V.; Scarpettini, A. F. Monitoring the Photothermal Reshaping of Individual Plasmonic Nanorods with Coherent Mechanical Oscillations. *J. Phys. Chem. C* **2018**, *122*, 29598–29606.

(42) Baffou, G.; Quidant, R. Thermo-plasmonics: using metallic nanostructures as nano-sources of heat. *Laser Photonics Rev.* **2013**, *7*, 171–187.

(43) Richardson, H. H.; Carlson, M. T.; Tandler, P. J.; Hernandez, P.; Govorov, A. O. Experimental and Theoretical Studies of Light-to-Heat Conversion and Collective Heating Effects in Metal Nanoparticle Solutions. *Nano Lett.* **2009**, *9*, 1139–1146.

(44) Mohammadi, M.; fazli, H.; karevan, M.; Davoodi, J. The glass transition temperature of PMMA: A molecular dynamics study and comparison of various determination methods. *Eur. Polym. J.* **2017**, *91*, 121–133.

(45) Dupaix, R. B.; Boyce, M. C. Constitutive modeling of the finite strain behavior of amorphous polymers in and above the glass transition. *Mech. Mater.* **2007**, *39*, 39–52.

(46) Startsev, O. V.; Lebedev, M. P. Glass-Transition Temperature and Characteristic Temperatures of α Transition in Amorphous Polymers Using the Example of Poly(methyl methacrylate). *Polym. Sci., Ser. A* **2018**, *60*, 911–923.

For Table of Contents Use Only:

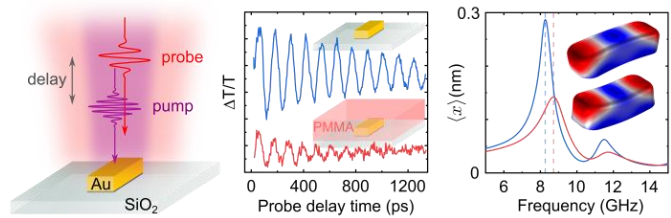

Determination of Nanoscale Mechanical Properties of Polymers via Plasmonic Nanoantennas

Hilario D. Boggiano, Rodrigo Berté, Alberto F. Scarpettini, Emiliano Cortés, Stefan A. Maier, and Andrea V. Bragas

Brief synopsis: A new method for determining the mechanical constants of thin polymer films is introduced, based on ultrafast optics techniques and numerical simulations. Using mechanical plasmonic nanoresonators that change their frequency of oscillation due to the presence of surrounding material, the values of the mechanical modules are extracted, and the value of the glass transition temperature is estimated.